\documentclass{PoS}

\title{Hadronic Freeze-Out in A+A Collisions meets the Lattice QCD Parton-Hadron Transition Line}

\ShortTitle{Hadronic Freeze-Out meets the Lattice QCD Parton-Hadron Transition Line}

\author{\speaker{R.~Stock}$^{1,2}$, F.~Becattini$^{3}$, M.~Bleicher$^{1}$, T.~Kollegger$^{1}$, T.~Schuster$^{4}$, J.~Steinheimer$^{5}$\\
        $^{1}$ Frankfurt Institute of Advanced Studies (FIAS), Frankfurt, Germany \\
        $^{2}$ Institut f\"{u}r Kernphysik, Universit\"{a}t Frankfurt, Germany \\
        $^{3}$ Universit\`{a} di Firenze and INFN Sezione di Firenze, Italy \\
        $^{4}$ Physics Department, Yale University, New Haven, CT, USA \\
        $^{5}$ Nuclear Science Division, Lawrence Berkeley National Laboratory, Berkeley, CA, USA \\
        E-mail: \email{stock@ikf.uni-frankfurt.de}}

\abstract{We analyze hadrochemical freeze-out in central Pb+Pb collisions at CERN SPS and LHC energies. Employing the UrQMD hybrid transport model we study the effects of the final hadron/resonance expansion phase on the hadron multiplicities established at hadronization. The bulk meson yields freeze out directly at hadronization whereas the baryon-antibaryon sector is subject to significant alterations, due to annihilation and regeneration processes. We quantify the latter changes by survival factors for each species which are applied to modify the statistical model predictions for the data. The modified SM analysis recovers the hadronization points, which coincide with the recent lattice QCD predictions of the parton-hadron transition line at finite baryochemical potential.}

\FullConference{8th International Workshop on Critical Point and Onset of Deconfinement,\\
		March 11 to 15, 2013\\
		Napa, California, USA}

\newcommand{\sqrts}{\ensuremath{\sqrt{s_{\textrm{\tiny NN}}}}}
\newcommand{\mub}{\ensuremath{\mu_{B}}}

\newcommand{\gev}{\ensuremath{\mathrm{\ GeV}}}

\newcommand{\mev}{\ensuremath{\mathrm{\ MeV}}}
\newcommand{\tev}{\ensuremath{\mathrm{\ TeV}}}
\newcommand{\fm}{\ensuremath{\mathrm{fm}}}

\begin{document}

\section{Introduction}

The first studies of meson production at the BEVALAC showed that the yield ratio e.g.\ of pions to participant baryons, once established during the high density phase of the collision, stayed constant throughout the expansion phase thus surviving all subsequent interactions. This observation was cast into the concept of hadro-chemistry, supposed to become stationary at a primordial ``hadrochemical freeze-out''~\cite{PRC1}, due to the rapid fall-off of inelastic transmutation, right at the beginning of the expansion phase. At its end, overall ``kinetic freeze out'' finally delivers the spectral observables, HBT etc.. This two freeze-out picture was subsequently, and tentatively, extended to all hadronic multiplicities, and it turned out that the yield distributions, over species, could be approximately understood to resemble grand canonical statistical equilibrium Gibbs ensembles~\cite{Becattini:2003wp,Andronic:2008gu,PRC3,BraunMunzinger2004:QGP3}, from AGS and SPS to RHIC energies.

The derived hadro-chemical freeze-out points, given in the parameters $(T, \mub)$ of the Gibbs ensemble, attracted attention as an early equilibrium snapshot of the dynamical evolution in A+A collisions. Moreover, as the experiment energies increased to top SPS and RHIC conditions, it was found that the statistical model (SM) temperatures saturated toward about $165\mev$ at small \mub. This coincided with the ``critical temperature'' of the parton-hadron phase transition as predicted by lattice QCD calculations~\cite{Karsch2004:QGP3} at zero \mub. Hadrochemical freeze-out thus appeared to coincide, closely, with QCD hadronization, locating the latter~\cite{Stock:1999hm}. This might have also explained the observation of statistical equilibrium of species, to result directly from the mechanisms of QCD hadronization~\cite{Heinz:1999kb,becareview,Stock:2008ru}, and not as an equilibrium limit of inelastic hadron interaction at and below $T = 165 \mev$. A similar equilibrium had been observed before, in elementary electron-positron annihilation to hadrons, and tentatively explained in the models of string hadronization~\cite{Andersson:1983ia} or cluster color neutralization~\cite{Webber:1983if}. The goal of experimentally locating the phase boundary in the QCD phase diagram~\cite{FodorAllton} appeared to be reached.

However, several subsequent observations have shown that this ideal picture of hadronization in central A+A collisions faced significant difficulties:
\begin{enumerate}
\item When statistical model analysis proceeded to finite baryochemical potential, at lower energies, it turned out that the freeze-out curve~\cite{Cleymans:2005xv} --- an interpolation of the freeze-out points in $(T, \mub)$ obtained at successive energies - disentangled from the lattice QCD prediction of the phase boundary~\cite{FodorAllton}, falling well below above $\mub = 300~\mathrm{MeV}$. What are we freezing-out from, here?
\item With the arrival of comprehensive, high accuracy data at many energies it was noticed that the SM fits resulted in unacceptably high $\chi^{2}$ per DOF, oftentimes yielding temperatures well below the former freeze-out curve. As a recent example, analysis of the ALICE central collision results at the LHC energy of $2.7 \tev$ resulted in recognition of an apparent ``non-thermal pion to proton/antiproton ratio''~\cite{Antinori:2011us}, the SM analysis~\cite{Andronic:2012dm} leading to temperatures well below the freeze-out curve as extrapolated to $\mub = 0$.
\item The effects of the final hadron/resonance expansion phase on the primordial hadronic yield distributions was analyzed in microscopic hadron transport models~\cite{Steinheimer:2012rd,Becattini:2012sq,Karpenko:2012yf} and in analytic blast-wave and Bjorken expansion models~\cite{Rapp:2000gy,Pan:2012ne}. It turned out that the inelastic sector that delivers the bulk meson output indeed freezes out immediately after hadronization, whereas baryon-antibaryon annihilation and regeneration processes continue throughout the hadronic expansion phase, inflicting changes of the primordial yields at the level of 20 to 50\%. This explains the unsatisfactory $\chi^{2}$ SM fits. There is no universal instantaneous freeze-out at hadronization. Furthermore, it was found that these annihilation effects generally lower the temperatures obtained in traditional SM analysis~\cite{Becattini:2012sq}.
\end{enumerate}

We shall illustrate the latter observations, from an UrQMD study of the expansion phase effects. Taking account of the yield changes in the baryon-antibaryon sector by applying UrQMD-derived ``survival factors'' for each species, as a modification of the SM analysis, we recover the primordial hadronization points~\cite{Becattini:2012sq,Becattini:2012xb}. Thus the initial goal of SM analysis can be met. Moreover, these hadronization points agree remarkably well with the recent predictions of lattice QCD for the parton-hadron phase boundary line~\cite{Endrodi:2011gv,Kaczmarek:2011zz} at finite \mub.

\section{UrQMD study of the hadronic expansion phase}

\begin{figure}
\begin{center}
  \includegraphics[width=0.45\columnwidth]{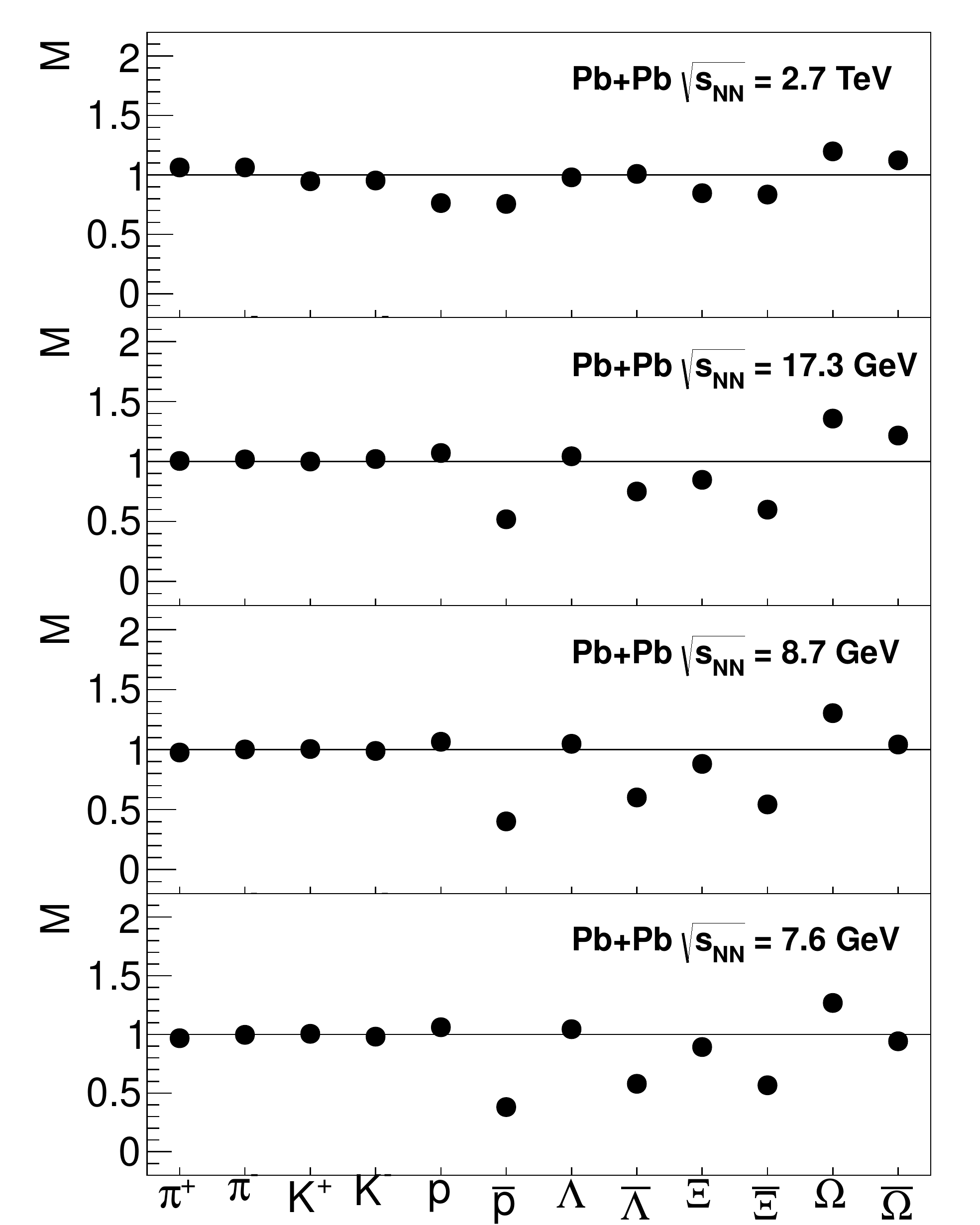} \\
  \includegraphics[width=0.45\columnwidth]{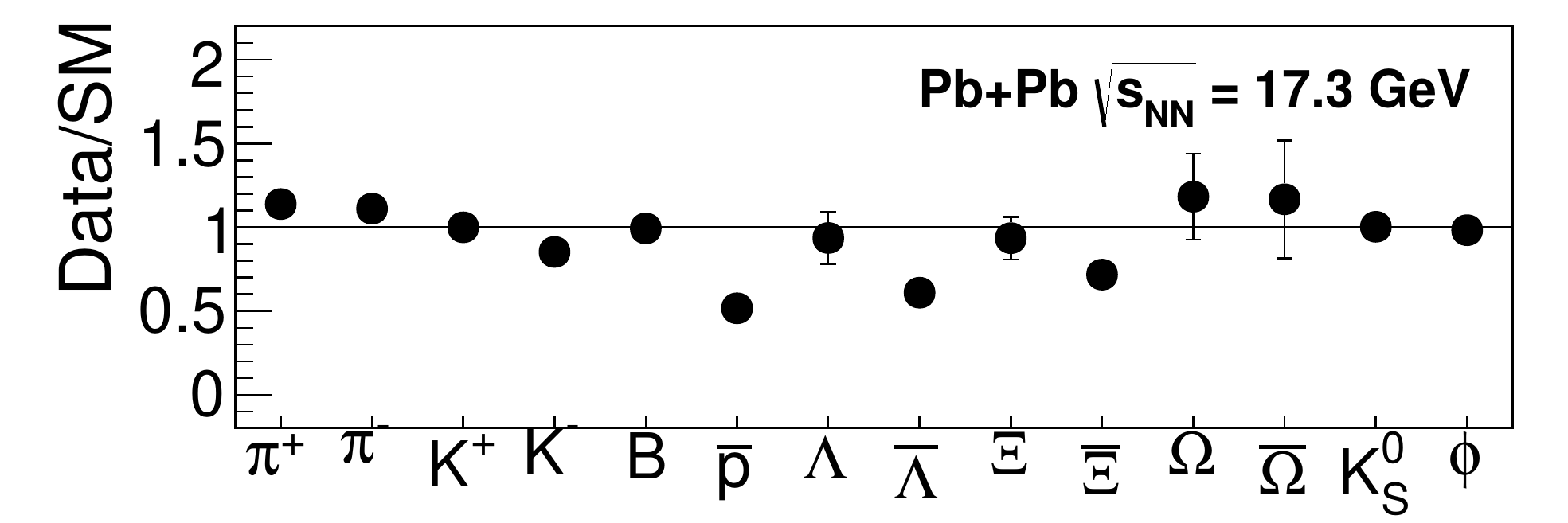}
  \caption{\label{fig:ModificationFactors} Top panel: Modification factors from UrQMD between particle multiplicities at hadronization and after the hadronic cascade afterburner stage for Pb-Pb collisions at $\sqrts = 2.7 \ \tev$, $17.3 \ \gev$, $8.7\ \gev$ and $7.6\ \gev$. Bottom panel: Ratio between central Pb-Pb collisions data at $\sqrts = 17.3 \gev$ and a statistical model fit excluding antibaryons~\cite{Becattini:2012sq}, for which the effect of the hadronic stage is largest.}
\end{center}
\end{figure}

We illustrate in Fig.~\ref{fig:ModificationFactors} the effects of the final hadron/resonance expansion phase on the observed yield distributions, as derived from the UrQMD hybrid version~\cite{Petersen:2008dd}. It features a 3+1 dimensional hydrodynamic expansion during the high density stage, terminated by the Cooper-Frye hadron formation mechanism. In order to account for the considerable time dilation that occurs toward large rapidity we have changed the ``isochronous'' procedure of~\cite{Petersen:2008dd}. We hadronize~\cite{Li:2008qm} in successive transverse slices, of thickness $\Delta(z)=0.2\ \fm$, whenever all fluid cells of that slice fall below a ``critical'' energy density, assumed here to be $0.8 \gev/ \fm^{3}$. We thus achieve a rapidity independent freeze-out temperature. The hadron distribution can be examined at this stage, emitting into vacuum. Alternatively, the UrQMD cascade expansion stage is attached to the Cooper-Frye output, as an ``afterburner''. The effect of this stage can be quantified by modification factors, for each hadron multiplicity, $M = N(\mathrm{Hydro+Aft})/N(\mathrm{Hydro})$. These factors are shown in Fig.~\ref{fig:ModificationFactors} (top panel). It illustrates the results obtained for central Pb+Pb collisions at SPS energies, $\sqrts=17.3 \gev, 8.7 \gev$
and $7.6 \gev$~\cite{Becattini:2012sq}, and for the present top LHC energy, $\sqrts=2.76\tev$~\cite{Steinheimer:2012rd}. At the SPS energies we see the bulk hadron output relatively unaffected by the afterburner, including the $\Xi$, $\Omega$ and $\bar \Omega$ yields. Whereas the other antibaryons, $\bar p$, $\bar \Lambda$ and $\bar \Xi$ are showing significant suppressions ranging from 50 to 25\%. At LHC energy the modification factors of baryons and antibaryons in Fig.~\ref{fig:ModificationFactors} become approximately equal as is to be expected in view of the particle-antiparticle symmetry prevailing here (with $\mub$ close to zero). The suppression pattern differs, in detail, from the pattern at SPS energy. It appears to be restricted to the $p$, $\bar p$, $\Xi$ and $\bar \Xi$ yields whereas the $\Lambda$, $\Omega$ and their antiparticles exhibit influences of a possible dynamical regeneration (see ref~\cite{Steinheimer:2012rd} for discussion).

Within the above model considerations the annihilation and/or regeneration effects inflict distortions of the initial equilibrium yield distributions
imprinted into the subsequent cascade phase by the grand canonical Cooper-Frye formalism. It is important to demonstrate that a quantitatively similar distortion pattern governs the experimentally observed hadron multiplicity data. To this end we have performed a SM analysis of the NA49 hadron yield data~\cite{Anticic:2010mp} for central Pb+Pb collisions at $17.3\gev$ excluding the most affected antibaryon species $\bar p$, $\bar \Lambda$ and $\bar \Xi$ from the fit procedure (see ref~\cite{Becattini:2012sq} for detail). The result is shown in the bottom panel of Fig.~\ref{fig:ModificationFactors}. It shows the ratios of data relative to SM predictions, for all species. We note that the bulk hadron data are well reproduced whereas the yields of $\bar p$, $\bar \Lambda$ and $\bar \Xi$ exhibit a strong suppression relative to the SM equilibrium distribution of multiplicities. The pattern quite closely resembles the UrQMD prediction in the upper panel of Fig.~\ref{fig:ModificationFactors}.

\begin{figure}[htbp]
\begin{center}
  \includegraphics[width=0.45\columnwidth]{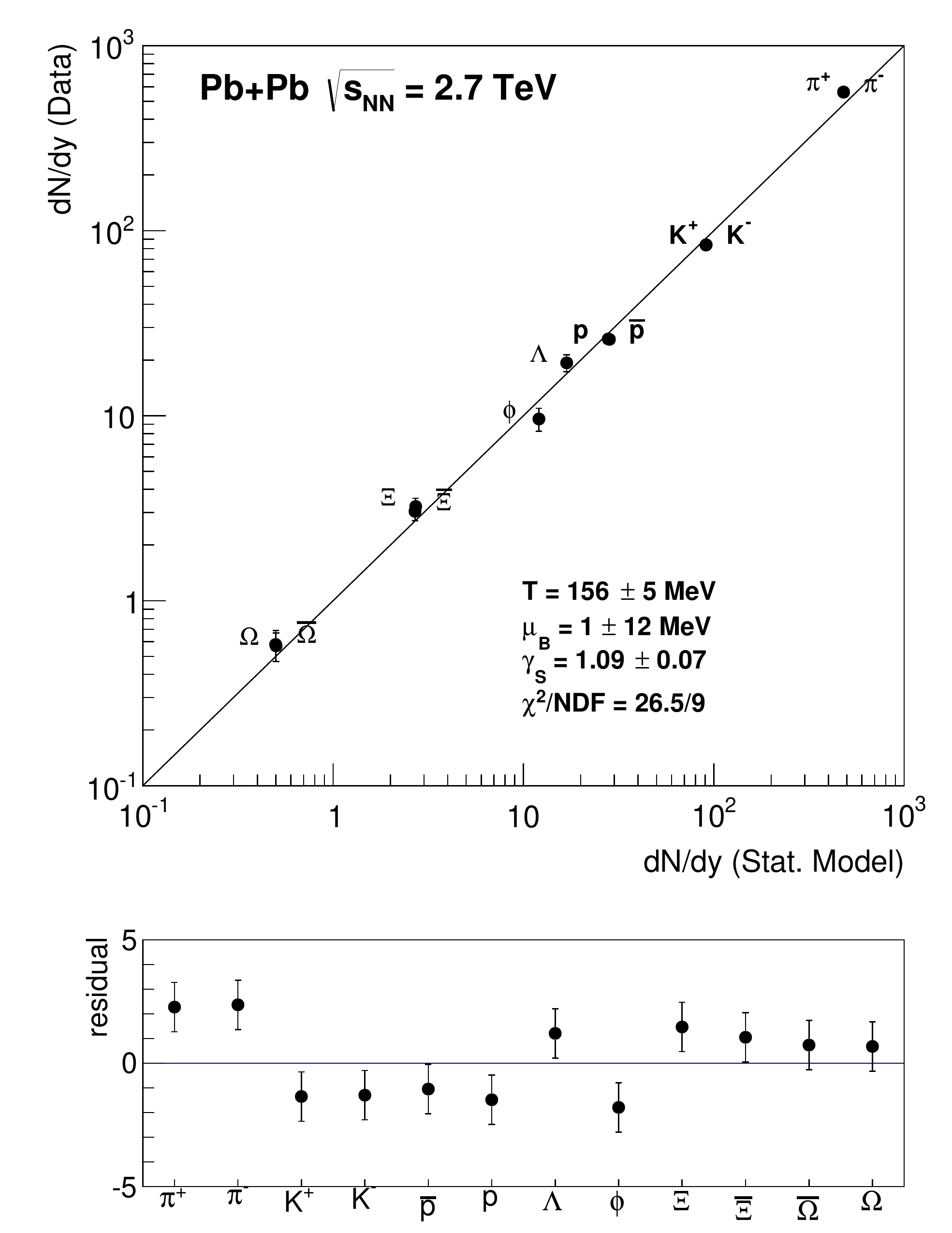}
  \includegraphics[width=0.45\columnwidth]{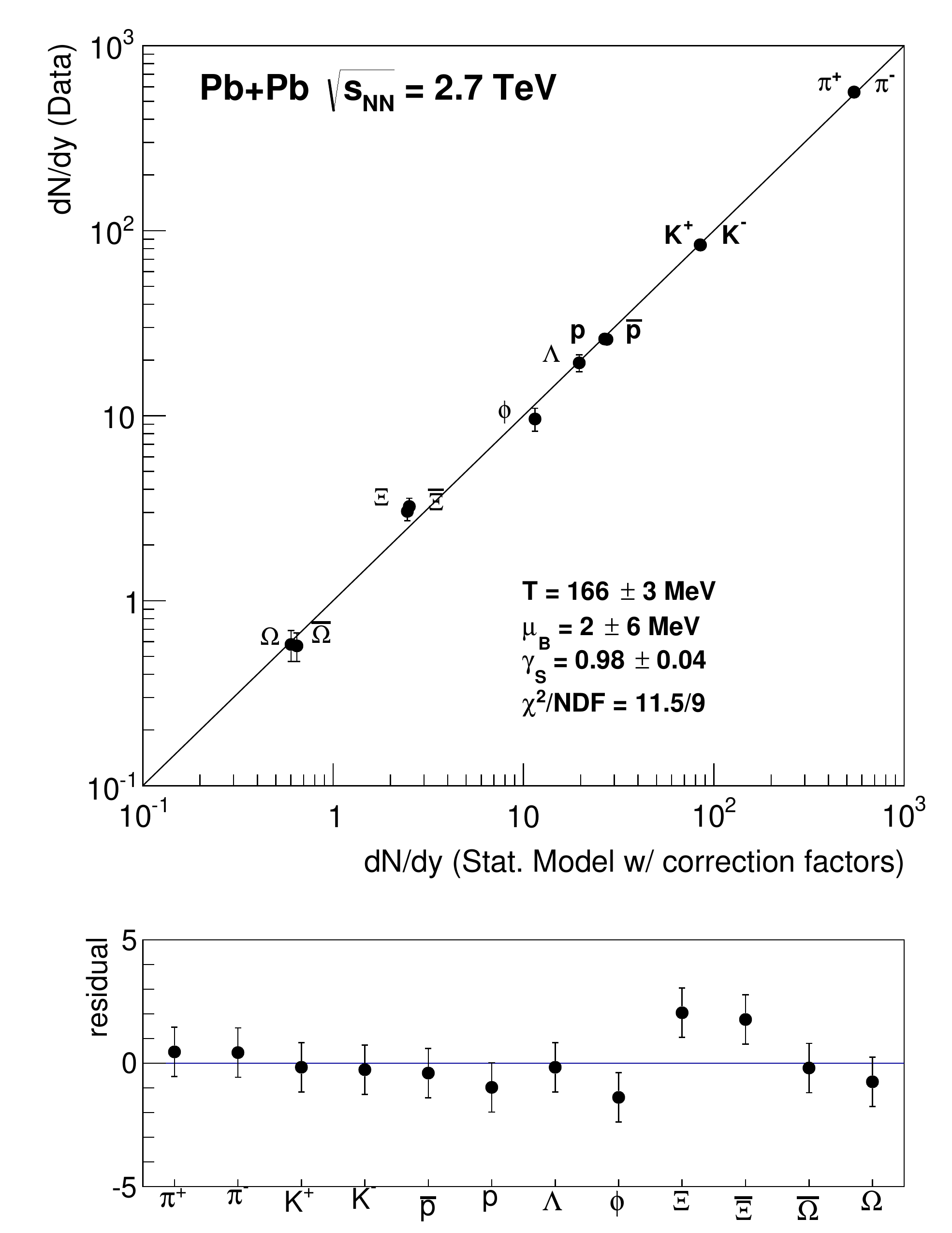}
  \caption{\label{fig:LHCFits} Statistical model fits to preliminary ALICE data~\cite{ALICE:2012QM} for $20\%$ central Pb-Pb collisions at $\sqrts = 2.7 \ \tev$ (left) and to the same data but with modification factors from UrQMD applied in the statistical model fits (right).}
\end{center}
\end{figure}

These observations lead to the idea to employ the UrQMD ``survival factors'' on face value: as modification factors employed in the SM analysis that aims at constructing a curve of last hadronic chemical equilibrium (denoted as LHCE). Such an analysis is shown in Fig.~\ref{fig:LHCFits}, applied to recent LHC ALICE data~\cite{ALICE:2012QM} for the 20\% most central Pb+Pb collisions at $2.76\tev$. The left panel gives the standard SM fit which is unsatisfactory, the bulk pion and kaon yields being missed at the cost of accounting for the baryon sector. Similar results have been obtained in ref.~\cite{Andronic:2012dm}. The right panel shows the analysis with  modification factors (the survival factors from UrQMD in Fig.~\ref{fig:ModificationFactors}) applied to the SM fit procedure. It yields a $(T,\mub)$ point at $(166 \mev,2 \mev)$, with improved $\chi^{2}$.

We repeated this analysis with the NA49 data for 5\% centrality selected Pb+Pb collisions~\cite{Anticic:2010mp} at $\sqrts = 7.6, 8.7$ and $17.3 \gev$. All obtained SM parameters are gathered in Tab.~\ref{tab:FitResults}. We do not include the RHIC data in the present analysis because the hadron multiplicities are not yet systematically corrected for feed-down into p, $\bar p$, $\Lambda$ and $\bar \Lambda$ from weak decays (an effect that counteracts the annihilation losses), and because a previous analysis~\cite{Manninen:2008mg} has met with considerable difficulties in cross-normalizing between the 3 experiments.  More data are forthcoming from the RHIC beam energy scan program~\cite{Mohanty:2011nm} which can be used to systematically extend the present SM analysis in the future.

\begin{table}
\begin{center}
 \begin{tabular}{c|c|c|c|c}
 & $T (\mev)$ & $\mub (\mev)$ & $\gamma_{S}$ & $\chi^{2}/NDF$ \\
 \hline
 \hline
 \multicolumn{5}{l}{Pb-Pb 20\% central $\sqrts = 2.7 \ \tev$} \\
 \hline
 Std. fit  & $ 156 \pm 5$ & $ 1 \pm 12$ & $ 1.09 \pm 0.07$ & $26.5/9$ \\
 Mod. fit  & $ 166 \pm 3$ & $ 2 \pm 6$ & $ 0.98 \pm 0.04 $ & $11.5/9$ \\
 \hline
 \hline
 \multicolumn{5}{l}{Pb-Pb 5\% central $\sqrts = 17.3 \ \gev$} \\
 \hline
 Std. fit  & $ 151 \pm 4$ & $ 266 \pm 9$ & $ 0.91 \pm 0.05$ & $26.9/11$ \\
 Mod. fit  & $ 163 \pm 4$ & $ 250 \pm 9$ & $ 0.83 \pm 0.04$ & $20.4/11$ \\
 \hline
 \hline
 \multicolumn{5}{l}{Pb-Pb 5\% central $\sqrts = 8.7 \ \gev$} \\
 \hline
 Std. fit  & $ 148 \pm 4$ & $ 385 \pm 11$ & $ 0.78 \pm 0.06$ & $17.9/9$ \\
 Mod. fit  & $ 161 \pm 6$ & $ 376 \pm 15$ & $ 0.72 \pm 0.06$ & $25.9/9$ \\
 \hline
 \hline
 \multicolumn{5}{l}{Pb-Pb 5\% central $\sqrts = 7.6 \ \gev$} \\
 \hline
 Std. fit  & $ 140 \pm 1$ & $ 437 \pm 5$ & $ 0.91 \pm 0.01$ & $22.4/7$ \\
 Mod. fit  & $ 156 \pm 5$ & $ 426 \pm 4$ & $ 0.81 \pm 0.00$ & $14.7/7$ \\
 \hline
 \hline
 \end{tabular}
 \caption{\label{tab:FitResults} Results of the statistical model fits to LHC and SPS data.}
\end{center}
\end{table}

\section{The QCD Phase Diagram revisited}

\begin{figure}
\begin{center}
  \includegraphics[width=0.6\columnwidth]{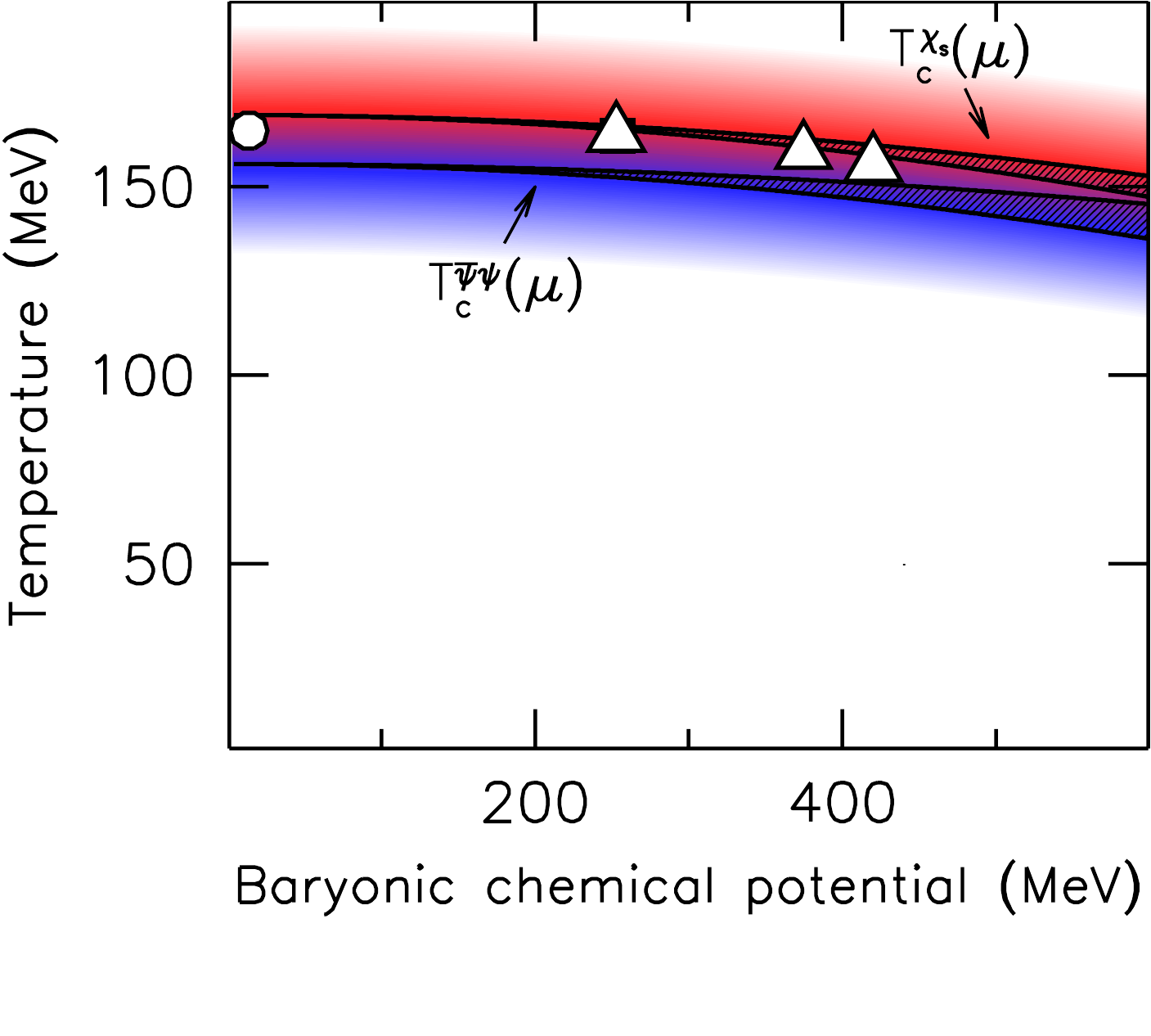}
  \caption{\label{fig:PhaseDiagram} Phase diagram of strongly interacting matter in the $\left( T, \mu_{B} \right)$ plane with predictions from lattice QCD calculations: the upper solid line is the critical temperature defined  through strange quark susceptibility, the lower one defined through the chiral condensate~\cite{Endrodi:2011gv, Kaczmarek:2011zz}. The colored areas represent the widths of the crossover transitions. The last hadronic chemical equilibrium (LHCE) points from our analysis are shown as an open circle (ALICE data) and open triangles (NA49 data), respectively. Figure adapted from~\cite{Endrodi:2011gv}.}
\end{center}
\end{figure}

Fig.~\ref{fig:PhaseDiagram} shows our principal result. The four obtained LHCE points are inserted into a phase diagram obtained recently by QCD lattice calculations at finite baryochemical potential, by Endrodi et al.~\cite{Endrodi:2011gv}. Similar results can be found in~\cite{Kaczmarek:2011zz}. The authors distinguish two different determinations of the position of the critical curve, based on the chiral condensate $\langle \bar \Psi \Psi \rangle$  and on the strange quark susceptibility $\chi_{S}/T^{2}$, respectively. The resulting two close curves $T_{\mathrm{C}} \left( \mub \right)$ cover baryochemical potentials up to $600\mev$. The LHCE points, determined above (which we identify with the hadronization points), follow the latter theoretical choice for the parton-hadron transition line. They are listed in Tab.~\ref{tab:FitResults}. One generally observes an upward shift of $T$ from chemical freeze-out (as analyzed in the ``conventional'' SM approach~\cite{Becattini:2003wp, Andronic:2008gu, BraunMunzinger2004:QGP3, Cleymans:2005xv, Andronic:2012dm}) to the LHCE points obtained upon application of the UrQMD modification factors. Except at $8.7 \gev$ it is accompanied by an improved $\chi^{2}$.

 A brief reflection is in order here concerning our employ of the UrQMD hybrid transport model~\cite{Petersen:2008dd}. Its account for the final hadron/resonance cascade evolution does not include the reverse of the annihilation processes, such as $p + \bar p$ to 5 pions, that could modify the survival factors~\cite{Rapp:2000gy, Pan:2012ne}. This is a general feature of all existing microscopic transport models. To account for such reverse reaction channels and establish, in principle, effects of detailed balance, one has to employ analytic models of fireball expansion~\cite{Rapp:2000gy, Pan:2012ne}. These, in turn, imply thermodynamically homogeneous collision volumes that miss the effects of local and surface density fluctuations. They thus overestimate both the annihilation and, in particular, the regeneration rates which scale with the fifth power of the density. Most remarkably, the main prediction~\cite{Rapp:2000gy, Pan:2012ne} is, again, a net loss of about 50\% in the $\bar p$ yield, similar to the UrQMD results shown in Fig.~\ref{fig:ModificationFactors}.

\section{Summary}

In summary we have shown that chemical freeze-out in A+A collisions does not occur synchronously for all hadronic species. It thus results in a yield distribution which deviates from an ideal Gibbs equilibrium situation, particularly in the baryon/antibaryon sector. We suggest to account for these effects of the final hadronic expansion phase by obtaining proper modification factors, for each species, from a microscopic model calculation, and applying them in the SM analysis. We have employed here the hybrid UrQMD model, and our SM analysis of LHC and SPS data recovers tentative points of last global chemical equilibrium which we identify with the QCD hadronization points, which is in line also with the observation that they fall on the parton-hadron coexistence line recently predicted by lattice QCD at finite baryochemical potential.

\section*{Acknowledgements}

We express our gratitude to the ALICE and NA49 Collaborations for providing their preliminary data. This work was supported by the Hessian LOEWE initiative through HIC for FAIR, by the Istituto Nazionale di Fisica Nucleare (INFN) and in part by the U.S. DOE under grant number DE-SC004168. We are also grateful to the Center for Scientific Computing (CSC) at Frankfurt and to the INFN Sezione di Firenze for providing the computing resources. JS and TS were supported in part by the Alexander von Humboldt Foundation as Feodor Lynen Fellows.

\end{document}